\documentclass[twoside]{dis09}
\usepackage[latin1]{inputenc}
\usepackage[dvips]{graphicx,epsfig,color}
\usepackage{wrapfig,rotating}
\usepackage{amssymb,amsmath,array}

\newcommand{\tw}{\textwidth}
\newcommand{\be}{\begin{equation}\nonumber}
\newcommand{\ee}{\end{equation}}
\newcommand{\bea}{\begin{eqnarray}}
\newcommand{\eea}{\end{eqnarray}}

\newcommand{\bc}{\begin{center}}
\newcommand{\ec}{\end{center}}

\newcommand{\bi}{\begin{itemize}}
\newcommand{\ei}{\end{itemize}}

\definecolor{red}{rgb}{1,0,0}
\definecolor{green}{rgb}{0.3,0.6,0.3}
\definecolor{blue}{rgb}{0,0,1}
\definecolor{darkgreen}{rgb}{0,0.39,0.00}

\begin{document}
\title{Parton distributions and small-x QCD at the Large Hadron Electron
Collider}

\author{Juan Rojo and Fabrizio Caola
%
%
\vspace{.3cm}\\
%
Dipartimento di Fisica, Universit\`a di Milano and\\
INFN, Sezione di Milano,\\ Via Celoria 16, I-20133 Milano, Italy\\
}

\maketitle

\begin{abstract}
The proposed Large Hadron Electron Collider (LHeC) at CERN would
bring Deep-Inelastic scattering into the unexplored TeV regime. 
The LHeC rich
physics program, among other topics,
 includes both precision SM measurements to complement
LHC physics as well as studies  of QCD
in the high energy limit.
 The present contribution reports on ongoing
studies within the NNPDF framework towards the LHeC CDR.
We study the impact
of LHeC simulated data on PDF uncertainties, in particular the small-$x$
gluon. We also assess the LHeC potential to disentangle between various
scenarios of small-$x$ QCD, including saturation models
and small-$x$ resummation. Finally, we explore how deviations
from DGLAP can be quantified in inclusive measurements.
\end{abstract}


\paragraph{Introduction}

The Large Hadron Electron Collider 
(LHeC)~\cite{Dainton:2006wd} is
a proposal for a Deep-Inelastic scattering facility in the TeV range
which would operate in parallel with the LHC. It would use the 7 TeV
LHC proton
beam colliding with a high energy electron beam, coming either from a
LHC-like ring or from a linear accelerator. The kinematical
coverage of such machine would extend
the HERA kinematical coverage by two orders of magnitude both
in $x$ and in $Q^2$.

After the experience at HERA, it is clear that the LHeC
physics potential includes the capability to probe the
nucleon structure and its flavour decomposition
with very high precision. However, the standard
approach to PDF determination~\cite{Nadolsky:2008zw,Martin:2009iq}
suffers from several shortcomings. The most important ones are
related to the fine-tuning of the PDF parametrizations and the statistical
definition of the associated PDF uncertainties
to the available dataset.

These shortcomings render difficult its application
to extrapolation regions like the LHeC kinematics.
In particular, in the standard PDF approach,
 PDF uncertainties are artificially reduced in 
extrapolation regions due to relatively simple
polynomial parametrizations employed,
thus making difficult a quantitative assessment of the
impact of new data from unexplored regions into the PDFs.
On top of that,
subtle deviations from DGLAP evolution which might
be present at small-$x$ are difficult to probe
with simple fixed
functional forms  because
 their lack of flexibility could lead to misleading results.

A method to bypass the above problems has been proposed
by the NNPDF collaboration. Within 
the NNPDF approach~\cite{f2ns,f2p,DelDebbio:2007ee,
Ball:2008by,Rojo:2008ke,nnpdf12}
(see also~\cite{Dittmar:2009ii}),
a combination of neural networks as universal unbiased interpolants with
Monte Carlo sampling of experimental data for error propagation render
the PDFs and associated uncertainties statistically faithful.\footnote{
The NNPDF methodology has also been applied to other physical
problems in \cite{Rojo:2006kr,tau,GonzalezGarcia:2006ay}.}

In this contribution we report on ongoing studies
of PDF determination and small-$x$ QCD within the
NNPDF approach. In particular, we concentrate on LHeC
pseudo-data at small-$x$. We consider 
 $F_2(x,Q^2)$ and $F_L(x,Q^2)$ simulated 
pseudo-data at small-$x$, in a scenario in which the LHeC
machine has
electron energy of $E_e=70$ GeV  and electron
acceptance of $\theta_e\le 179^o$, for an integrated luminosity of
 $\int\mathcal{L}=1$ fb$^{-1}$. Full NC and CC cross-section
simulated pseudo-data
in various other machine scenarios are available, and their are
under current scrutiny.

\begin{figure}
\begin{center}
\includegraphics[width=0.75\tw]{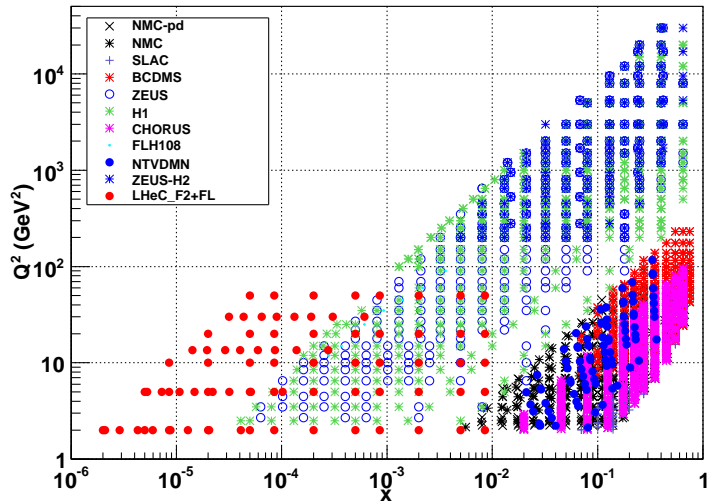}
\end{center}
\caption{\small The kinematical coverage of the
LHeC pseudo-data used in the present studies, together
with the data already included in the 
reference NNPDF1.2 dataset. \label{fig:kin}}
\end{figure}

The reference baseline for the studies presented in this
contribution is the recent NNPDF1.2 parton set~\cite{nnpdf12}, a PDF analysis
of all relevant inclusive DIS data together with
 neutrino charm production
to constrain strangeness. The kinematics of the pseudo-data,
together with that of the NNPDF1.2 analysis are shown 
in Fig.~\ref{fig:kin}. The average total uncertainty of the simulated
$F_2$ pseudo-data is $\sim 2\%$, while that of 
$F_L$ is $\sim 8\%$.

\paragraph{Constraining parton distributions at small-$x$}
In spite of the wealth of precision data on small-$x$ structure
functions at HERA, some PDFs, most notably the gluon, have still
rather large uncertainties in this region~\cite{Ball:2008by}. 
In order to quantify
how these PDF uncertainties would be reduced with LHeC data,
we have repeated the NNPDF1.2 analysis with the
addition of the LHeC pseudo-data, with central values
from the NNPDF1.0 predictions and 
experimental uncertainties corresponding to the
simulated LHeC scenario described above. The joint
data set is shown in Fig.~\ref{fig:kin}.

\begin{figure}
\begin{center}
\includegraphics[width=0.49\tw]{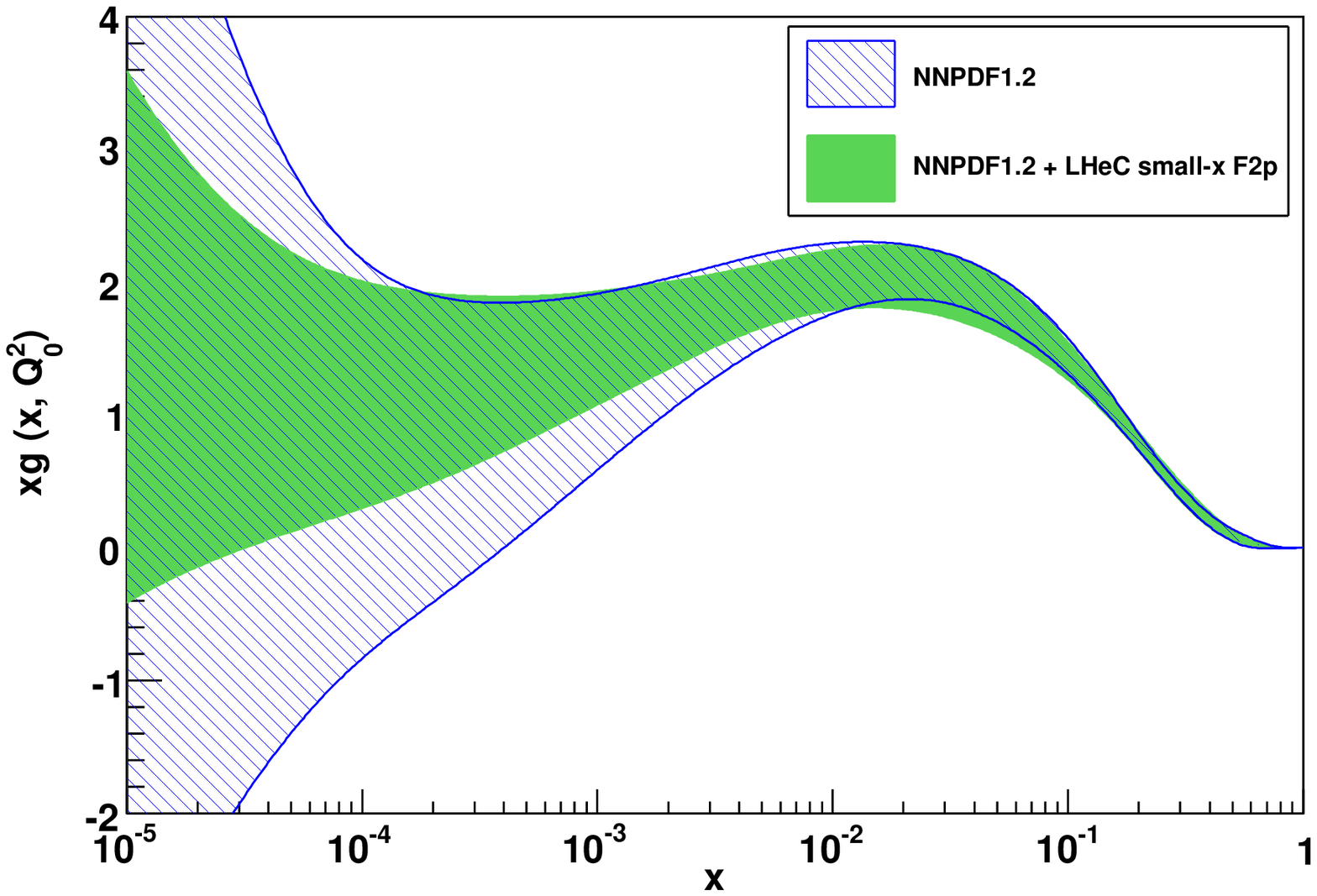}
\includegraphics[width=0.49\tw]{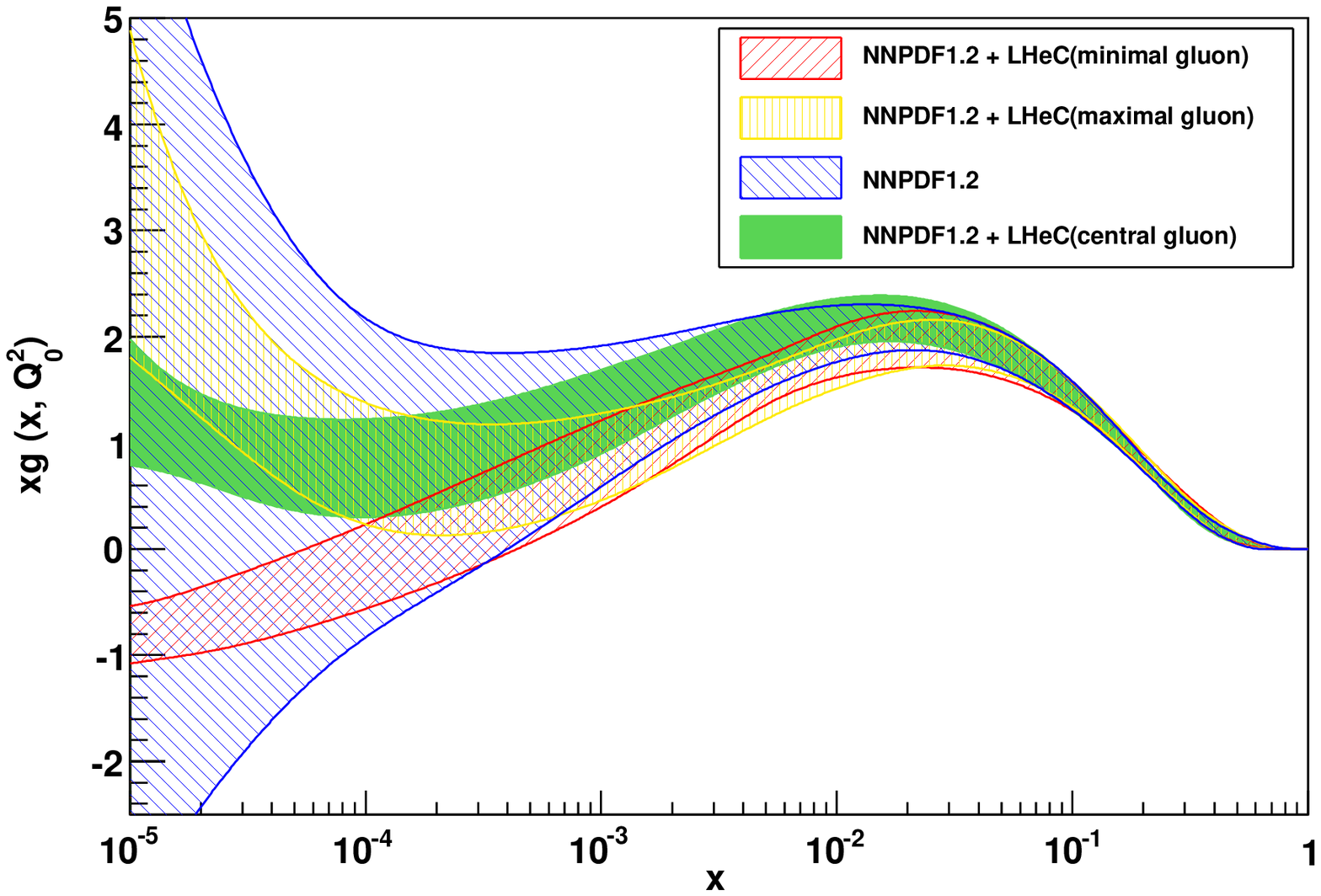}
\end{center}
\caption{\small Left: the gluon which results from a fit
which includes on top
of the NNPDF1.2 dataset \cite{nnpdf12} 
only $F_2$ LHeC pseudo-data. Right: the gluon PDF determined from
the same dataset 
together with $F_2$ and $F_L$ LHeC pseudo-data, generated from the
central and $\pm$1-$\sigma$ NNPDF1.0 gluons.
\label{fig:gluon-extremes}}
\end{figure}

First of all, we include only $F_2$ LHeC
pseudo-data into the analysis. Although the fit is as expected 
perfect, the reduction of the small-$x$ uncertainties is rather
moderate, as shown in Fig.~\ref{fig:gluon-extremes}
(left). Our results therefore indicate
that only $F_2$ data, even if very accurate,
  is not enough to pin down the gluon at small-$x$,
due to the fact that the gluon
PDF only enters through scaling violations and higher order corrections.

The next step consists of  the
addition of the complete $F_2$ and $F_L$ pseudo-data.
In this case, the joint fit with $F_2$ and $F_L$ pseudo-data
leads to a sizable decrease of the small-$x$ gluon uncertainties,
which can be understood from the greater sensitivity of $F_L$
to the gluon PDF.
To quantify more these results, 
we have generated LHeC pseudo-data in three different scenarios: one
where the gluon is the central NNPDF1.0 gluon and two more were the
gluon sits near the 
associated $\pm 1$-$\sigma$ envelope. We have repeated the joint
$F_2+F_L$ analysis
in these 
three cases: results are shown in Fig.~\ref{fig:gluon-extremes} (right).
As expected, after the fit the three extreme scenarios for the 
small-$x$ gluon can be precisely disentangled.
Therefore, it is clear
 that the LHeC has the potential to pin down with great precision
the behaviour of the low-$x$ gluon, but only after accurate measurements of
$F_L$ are performed.

\paragraph{Probing small-x QCD}
The LHeC would also provide us with an improvement
of our understanding of the small-$x$ dynamics of QCD.
Indeed, even after years of intensive study at HERA, no convincing
evidences for departures from standard DGLAP evolution
has been found. For example, geometric scaling of HERA data,
which was thought to provide a clear signal for saturation, was
recently shown to be consistent with linear QCD
evolution as well~\cite{Caola:2008xr}. 
The situation could be different at the
LHeC, with its extended kinematical coverage at small-$x$
(see Fig.~\ref{fig:kin}).

In order to test whether or not a DGLAP analysis can 
reproduce theoretical predictions
which deviate from pure DGLAP in inclusive measurements, 
LHeC pseudo-data has
been generated not within the DGLAP framework, as 
in the previous sections, but rather from two
different models: the AAMS09 model~\cite{Albacete:2009fh}, which is based
on BK evolution with running coupling, and the 
FS04 model~\cite{Forshaw:2004vv}, based
on the dipole model.

We have repeated the PDF analysis of the previous section but with these
new pseudo-data. Although clearly the procedure is not
consistent (for example, PDF error reduction would be meaningless
in this case), it provides an illustration of a potential analysis
technique which ultimately should be applied to experimental data.
For both the AAMS09 and the FS04 models the conclusions of the
study are the same: the DGLAP analysis
reproduces perfectly the $F_2(x,Q^2)$ pseudo-data,
which implies that although
the underlying physical theories are different,
from a practical point of view the small-$x$ extrapolations
of AAMS09 and FS04 for $F_2$ are rather similar
to DGLAP-based extrapolations. 

\begin{figure}
\begin{center}
\includegraphics[width=0.99\tw]{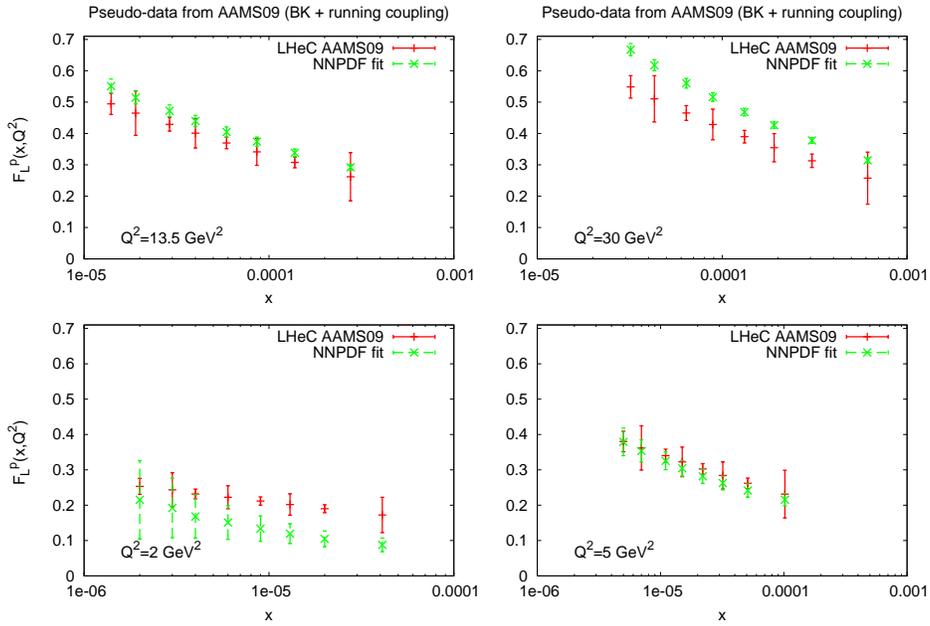}
\end{center}
\caption{\small The results of the 
combined DGLAP analysis of the NNPDF1.2 data set and the
LHeC pseudo-data for $F_L(x,Q^2)$ in various 
$Q^2$ bins generated with the AAMS09 model. \label{aams}}
\end{figure}

The situation however is different
for $F_L(x,Q^2)$: provided the level arm in $Q^2$ is large enough,
the DGLAP analysis fails to reproduce simultaneously $F_L$ in all the
$Q^2$ bins, and thus the overall $\chi^2$ is very large, a clear
signal of the departure from fixed order DGLAP of the simulated
pseudo-data. This effect is illustrated in Fig.~\ref{aams}, where
the results of the DGLAP analysis are compared with the 
LHeC pseudo-data generated from the AAMS09 model.

There exists however other scenarios for QCD at small-$x$ than
saturation/dipole models.
In particular, 
linear QCD evolution with resummation of BFKL small-$x$ logarithms
is the natural extension of standard DGLAP evolution.
Recently, the full set of small-$x$ resummed splitting functions
and coefficient functions became available~\cite{Altarelli:2008aj}.
The results of Ref.~\cite{Altarelli:2008aj} have been used
to compute resummed K-factors~\cite{AltarelliPrel}, 
defined as the ratio of structure
functions NLO small-$x$ resummed over fixed
order NLO, as a function of
$(x,Q^2)$. These K-factors can be used for realistic, though qualitative,
phenomenological studies of the impact of small-$x$ resummation.

\begin{figure}
\begin{center}
\includegraphics[width=0.99\tw]{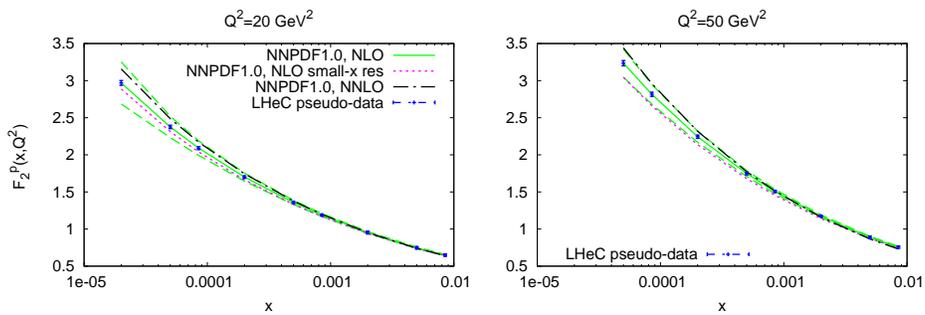}
\end{center}
\caption{\small A comparison of various
approximations to linear low-$x$ QCD for $F_2$
at the LHeC: the NNPDF1.0
prediction which includes PDF
uncertainties (green lines) and the
NNPDF1.0 result corrected with the
NNLO (black, dot-dashed) and NLOres (violet, short-dashed)
K-factors. The expected 
 experimental precision at the LHeC
is also shown for illustration. \label{f2resum}}
\end{figure}

We have used these resummed
K-factors to estimate the feasibility of the
LHeC to disentangle between scenarios for small-$x$ linear
QCD: NLO, NNLO and NLOres. 
In Fig.~\ref{f2resum} we show the LHeC pseudo data
for $F_2(x,Q^2)$ at small-$x$ compared with the NLO NNPDF1.0
prediction (including the associated
PDF uncertainties) and the corresponding NNLO and NLOres
computations, obtained from the NLO one with
these K-factors. 
Fig.~\ref{f2resum} seems
to indicate that a PDF analysis capable of implementing both the
the NNLO and NLOres computations
of physical observables has the potential to disentangle between these
two scenarios of small-$x$ QCD, given the foreseen experimental
accuracy at the LHeC.

\paragraph{Departures from DGLAP}
It is clear from the previous discussion that there is
some contradiction between the two goals of our study: either
we determine the PDFs or we
 find evidence for saturation or resummation. However,
both these goals require the same first step: we have to determine
the kinematic region where saturation/resummation effects,
or more general, departures from fixed-order
DGLAP evolution, start to
play a role, if any.

The idea is therefore to single out a {\it safe} region, where the
standard PDFs extraction via fixed order
 DGLAP is reliable, and a {\it small-$x$}
region where deviations from pure fixed order DGLAP could provide evidence
for saturation or resummation. The determination of these kinematic 
regions is a highly non trivial task: both BFKL and non-linear 
effects are known to be rather moderate
in the HERA region, and thus are
 difficult to observe in inclusive observables. In particular,
 they could be absorbed in the initial condition for flexible enough
parametrizations of the PDFs. This might already be the case at HERA
for $F_2$, and if so even more at the LHeC. 

A possible approach to this problem is the following. First we repeat
the global PDF
analysis removing subsets of data where small-$x$ effects could play some
role. Then we determine whether NLO DGLAP is able to reproduce the excluded 
data or not. A tension between the actual data and the DGLAP prediction
should mark the onset of some saturation/resummation effect. As a cross
check, we can assess the NLO DGLAP fit quality in the fitted data region: 
the fit quality should improve if there is some tension between DGLAP and
the actual data in the excluded region. Note that deviations from DGLAP
are known to be rather moderate,
hence our approach is meaningful only on statistical grounds. It is therefore
mandatory to perform a PDF analysis with no parametrization bias and with
faithful uncertainty estimation~\cite{Ball:2008by}. 
Related studies of the stability of global analysis
within the standard PDF approach 
have been reported in~\cite{Martin:2003sk,Huston:2005jm}.

\begin{figure}
\begin{center}
\includegraphics[width=0.49\tw]{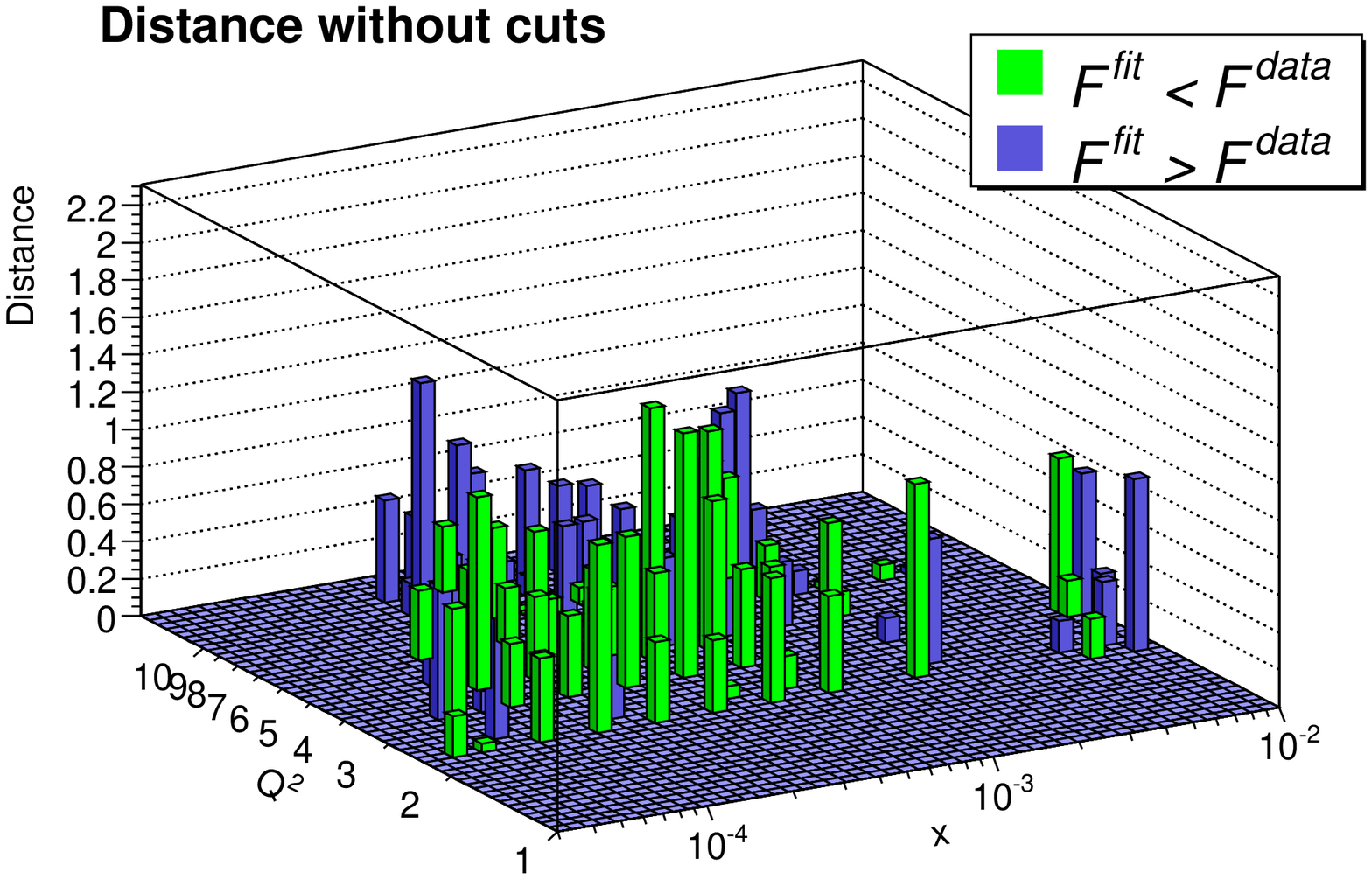}
\includegraphics[width=0.49\tw]{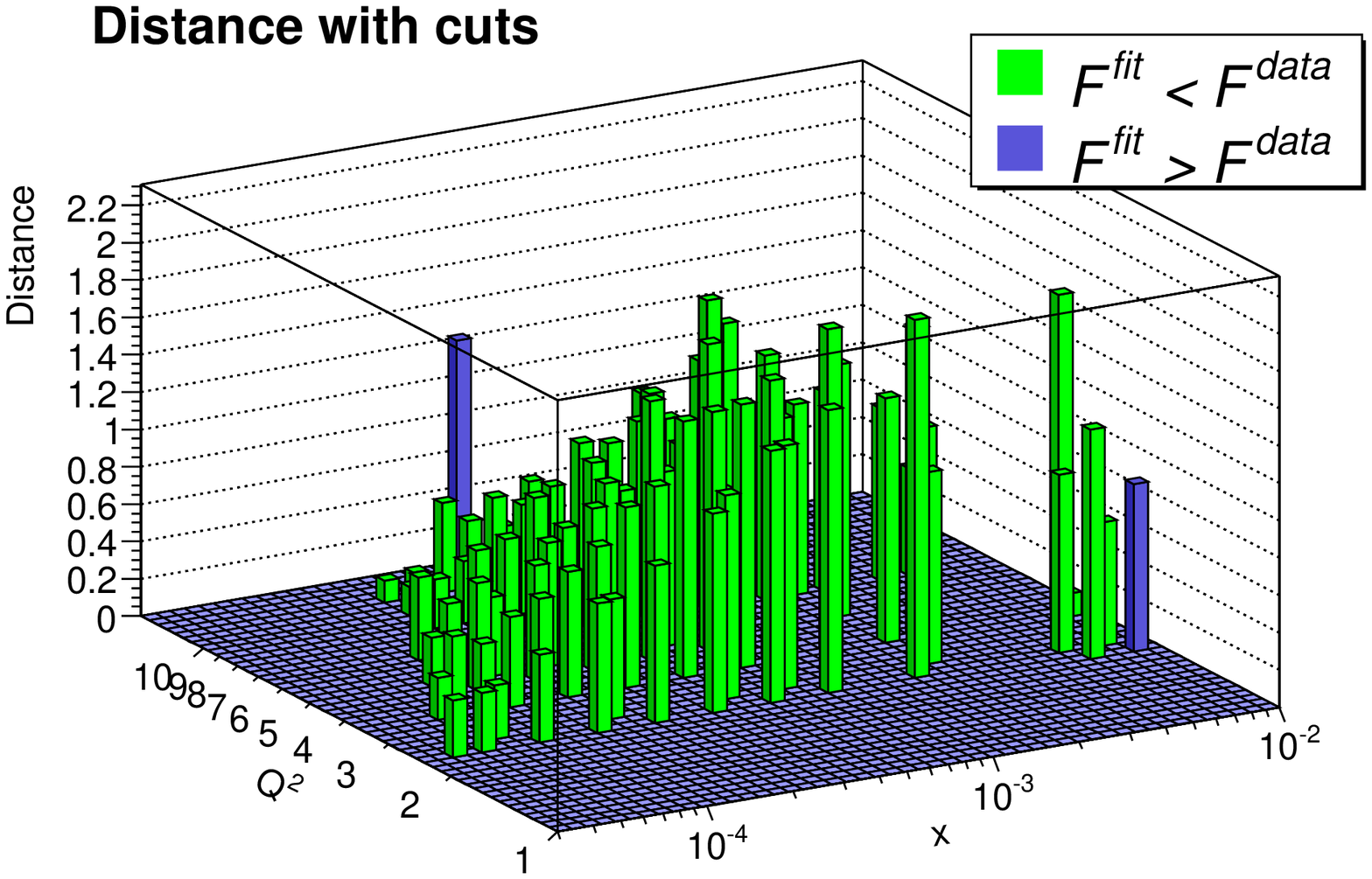}
\end{center}
\caption{\small Left: distances, Eq. \ref{eq:dist},
 in the $Q^2< 1.5~x^{-0.3}$ HERA region computed
using a global fit with these data points included. 
Right: distances in the same region computed when 
these points are excluded from the global fit. 
\label{fig:distance}}
\end{figure}

This approach has been
 applied to search for DGLAP deviations in the small-$x$ HERA
data. Taking as a reference the NNPDF1.2 analysis \cite{nnpdf12}, 
we excluded data points with a
saturation-inspired cut $Q^2\ge Q_s^2(x)\equiv A x^{-0.3}$, 
with $A$ ranging from $0.2$ to $1.5$. In order to 
quantify deviations from DGLAP we computed the distance $d(x,Q^2)$ between 
the DGLAP extrapolation $F^{\rm fit}$ for an 
observable $F$ and the actual data 
$F^{\rm data}$, defined as
\begin{equation}
d(x,Q^2)=\sqrt{\frac{F^{\rm fit}(x,Q^2)-F^{\rm data}(x,Q^2)}{\sigma_{\rm fit}^2+\sigma_{\rm data}^2}}\times \text{sign}\left(F^{\rm data}-F^{\rm fit} \right) \ .  \label{eq:dist}
\end{equation}
A typical result for the cut $Q^2\ge 1.5~ x^{-0.3}$ is shown in Fig.~\ref{fig:distance}.
Note that while in the global fit distances seem uncorrelated, in the 
 fit with the kinematical cut
$Q^2\ge 1.5~ x^{-0.3}$ there seems to be a hint of a correlation, that is, 
the NLO DGLAP prediction tends to be smaller than actual data. 
A systematic study is in progress using these
methods in order to determine the
statistical significance, if any, of departures from
DGLAP in inclusive small-$x$ data. The ultimate validation
of the method will be its application to LHeC pseudo-data, where
there the underlying physics can be varied within various
scenarios.

\paragraph{Outlook}
This contribution summarizes some of the studies performed
within the NNPDF framework in order to assess
the physics potential of the LHeC
as a probe of the nucleon structure and of small-$x$ QCD
dynamics. From these preliminary studies, one
solid conclusion is that the importance
of accurate measurements of $F_L(x,Q^2)$ should
be emphasized. Ongoing work towards the LHeC 
Conceptual Design Report includes
the generalization of the PDF analysis to the complete LHeC
data set for various scenarios and the impact of the reduction
in PDF uncertainties on LHC phenomenology.

\paragraph{Acknowledgments}
It is a pleasure  to
 acknowledge useful discussions with  J.~L.~Albacete, N.~Armesto, S.~Forte and
P.~Newman. J.R. acknowledges the hospitality
of the CERN TH Division where part of this work was completed.

\begin{footnotesize}




\end{footnotesize}


\end{document}